\documentclass[prl,aps,floats,superscriptaddress,amsfonts,twocolumn,showpacs]{revtex4}

\usepackage{graphicx}

\begin{document}

\newcommand{\ket} [1] {\vert #1 \rangle}
\newcommand{\bra} [1] {\langle #1 \vert}
\newcommand{\braket}[2]{\langle #1 | #2 \rangle}
\newcommand{\proj}[1]{\ket{#1}\bra{#1}}
\newcommand{\mean}[1]{\langle #1 \rangle}
\newcommand{\opnorm}[1]{|\!|\!|#1|\!|\!|_2}

\title{Unconditional optimality of Gaussian attacks against continuous-variable QKD}

\author{Ra\'{u}l Garc\'{\i}a-Patr\'{o}n}
\affiliation{QuIC, Ecole Polytechnique, CP 165,
Universit\'e Libre de Bruxelles, 1050 Bruxelles, Belgium}

\author{Nicolas J. Cerf}
\affiliation{QuIC, Ecole Polytechnique, CP 165,
Universit\'e Libre de Bruxelles, 1050 Bruxelles, Belgium}

\begin{abstract}
A fully general approach to the security analysis of continuous-variable 
quantum key distribution (CV-QKD) is presented. Provided that the quantum
channel is estimated via the covariance matrix of the quadratures,
Gaussian attacks are shown to be optimal against
all eavesdropping strategies, including collective and coherent attacks.
The proof is made strikingly simple by combining a physical model of measurement, an entanglement-based description of CV-QKD,
and a recent powerful result on the extremality of Gaussian states
[Phys. Rev. Lett. 96, 080502 (2006)].
\end{abstract}


\maketitle

Continuous-variables quantum information \cite{CVBook} 
has attracted a rapidly increasing interest over the past few years.
Several QKD schemes based on a Gaussian modulation of 
coherent states of light combined with homodyne 
or heterodyne detection have been proposed \cite{Gross02,Ralph04} and 
experimentally demonstrated \cite{Nature03,Jerome05}.
These protocols have the advantage of being based on standard optical telecom components and thereby of working at high repetition rates compared to the schemes based on single-photon detectors. The first security proof of CV-QKD was restricted to Gaussian individuals attacks \cite{Cerf01,Gross02,Nature03,Ralph04}. In such an attack, the eavesdropper (Eve) is assumed to interact individually - according to a Gaussian map - with each of the signal pulses sent over the line, and then to perform a Gaussian (homodyne or heterodyne) measurement on her probe after the basis information (if any) is disclosed but before the full classical post-processing. Later on, it was shown that non-Gaussian individual attacks cannot beat Gaussian attacks \cite{Cerf04}, so that studying the security against Gaussian individual attacks is quite justified. This proof extends to the case where Eve attacks finite-size blocks of pulses, but does not cover the important class of collective attacks, where Eve jointly measures all her probes (each having interacted with a signal pulse) after the classical post-processing has taken place \cite{Renato05,RenatoPhD,Winter03}. The security versus Gaussian collective attacks was recently studied in \cite{Gross05,Navascues05}, but a definitive proof of the optimality of Gaussian attacks was missing.

In this Letter, we prove that the optimal collective attack reduces to a Gaussian attack that is completely characterized by the covariance matrix of the quadratures observed by the emitter (Alice) and receiver (Bob). This optimality is probably even stronger in view of the recent result that the most general attacks, namely coherent attacks (where Eve coherently interacts with all signal pulses and performs a joint measurement after the classical post-processing), 
cannot outperform collective attacks \cite{Renato05,RenatoPhD},
implying that it is sufficient to check the security of QKD against collective attacks.

\textit{One-way QKD protocols with Gaussian continuous variables}
are divided in two steps, a quantum communication part followed by a classical post-processing. In the quantum part, Alice sends either a displaced squeezed state 
encoding a random Gaussian variable or a coherent state encoding two
Gaussian variables. Then, Bob performs either homodyne (active basis-choice) or heterodyne measurement (no basis-choice) on the received states
(not necessarily Gaussian) in order to decode Alice's variable. 
Once Alice and Bob have collected a sufficiently large list of correlated data,
they proceed with the classical post-processing.
Unless Alice sent coherent states and Bob did heterodyne measurement, they first apply a \textit{sifting}, where they compare the chosen encoding and measurement quadratures ($x$ or $p$) and keep only the values for which the quadratures match. 
Then, they apply \textit{parameter estimation}, i.e., they calculate the covariance matrix $\gamma_{AB}$ of their correlated variables 
from a randomly chosen sample of their data.
The optimal attack being Gaussian (as we will prove below), $\gamma_{AB}$  completely characterizes the channel as the first-order 
moments of the quadratures do not play any role. Finally,
they apply one-way \textit{error correction} and \textit{privacy amplification} to distill a secret key. The error correction can be done in two ways: either direct reconciliation (DR), where Bob corrects his data to Alice's ones, or reverse reconciliation (RR), where Alice's and Bob's roles are interchanged
\cite{Nature03}.

\begin{figure}[!t!]
\begin{center}
\includegraphics[width=5cm]{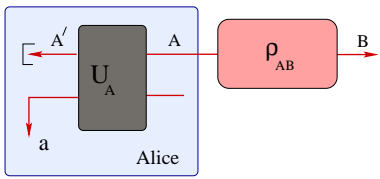}
\end{center}
\caption{Alice's measurement of system $A$ of the bipartite state $\rho_{AB}$, giving the result $a$. Equivalently, $a$ denotes the internal state of a preparer who prepares system $B$ according to $a$.}
\label{measure}
\end{figure}

\textit{Physical model of measurement}.
Assume Alice and Bob share a quantum state $\rho_{AB}$ and Alice then makes a von Neumann measurement on system $A$, obtaining the outcome $a$ distributed according to the probability distribution $p(a)$. This measurement can be realized by applying an appropriate unitary operation $U_{A}$ on $A$ together with an ancilla, and subsequently observing the state of this ancilla while tracing over the resulting quantum system $A'$ (see Fig.~\ref{measure}).
Considering the ancilla as a physical system, noted as $a$ after the action of $U_A$, the joint state of $a$ and $B$ after the measurement is
\begin{equation}
\rho_{aB}=\int da \; p(a)\proj{a}\otimes\rho_{B}^{a}.
\label{prepXB}
\end{equation}
Given the block-diagonal structure of $\rho_{aB}$,
the quantum mutual entropy $S(a{\rm:}B)$ can be shown to coincide with the Holevo bound $\chi_{aB}=S(\rho_B)-\int da \; p(a) S(\rho_B^a)$ \cite{Cerf96H}. Note that the situation here is fully equivalent to that where $a$ is a classical preparer and $B$ is a quantum preparation. Now, assume Bob measures his system $B$ by means of the unitary $U_B$ in a similar way as Alice. The
resulting joint state is given by the diagonal density operator,
\begin{equation}
\rho_{ab}=\int da \;db \;p(a,b)\proj{a}\otimes\proj{b}.
\label{prepXY}
\end{equation}
The quantum mutual entropy $S(a{\rm:}b)$ then simply reduces to the Shannon mutual information $I_{ab}$ between the preparer's and the measurer's internal states. The Holevo bound on the accessible information then becomes a straightforward consequence of the strong subadditivity of von Neumann entropies, namely \cite{Cerf96H}
\begin{equation}
I_{ab}=S(a{\rm:}b)\leq S(a{\rm:}bB')=S(a{\rm:}B)=\chi_{aB}.
\label{HolTh}
\end{equation}

\begin{figure}[!t!]
\begin{center}
\includegraphics[width=8cm]{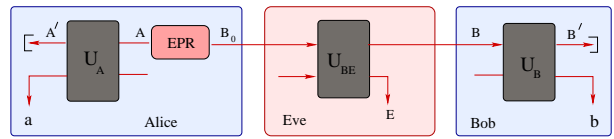}
\end{center}
\caption{Entanglement-based scheme for CV-QKD. Alice's preparation is modelled by a measurement $U_A$ on her half of an EPR pair. The channel is modelled by an unitary interaction between mode $B$ and Eve ancilla's $E$. Finally, Bob's measurement is modelled by $U_B$.}
\label{entCVQKD}
\end{figure}

\textit{Entanglement-based version of CV-QKD.}
The description of any prepare-and-measure CV-QKD protocol using its equivalent entanglement-based scheme is very convenient for security analyses \cite{Gross03}. Indeed, all protocols based on the Gaussian modulation of Gaussian states and homodyne (or heterodyne) measurement can be described in a unified way, see Fig.~\ref{entCVQKD}. Alice and Bob are assumed to share a bipartite quantum state $\rho_{AB}$, whose purification is given to Eve. Alice's measurement of $A$ is equivalent to a preparation scheme where she randomly chooses $a$, according to $p(a)$, and sends the state $\rho_{B_{0}}^{a}$ in the quantum channel so that Bob receives the state $\rho_{B}^{a}$ at the output. The unitary $U_A$ determines which measurement is performed: homodyne measurements, corresponding to the preparation of squeezed states, or heterodyne measurements, corresponding to the preparation of coherent states ($a$ then collectively denotes two real numbers).
The maximal information that is accessible to Bob is given, in principle, by 
$\chi_{aB}=S(a{\rm:}B)$. In practice, however,  
Bob applies an homodyne (or heterodyne) measurement on $B$, giving $b$, so the actually extracted information is $I_{ab}=S(a{\rm:}b)$. Since there are two possible encodings at Alice's station and two possible measurements at Bob's station, there exist four Gaussian protocols (three of them having been described in \cite{Gross02,Ralph04,Cerf01}).

Consider now that Eve performs a collective attack: she interacts individually with each signal pulse sent by Alice, stores her resulting probes 
in a quantum memory, and then applies a joint measurement over them at the end of the classical post-processing. As shown in \cite{Renato05,RenatoPhD}, her information is then limited by the Holevo bound $\chi_{aE}=S(\rho_E)-\int da\; p(a)S(\rho_E^a)$. 
Because Eve holds the purification of $\rho_{AB}$,
this bound can be calculated from $\rho_{AB}$: for example, when Alice and Bob apply the same measurement, it reads $\chi_{aE}=S(\rho_{AB})-\int da\; p(a)S(\rho_B^a)$.
If $\rho_{AB}$ is assumed to be Gaussian, then $\chi_{aE}$ can be directly computed from $\gamma_{AB}$ \cite{Gross05,Navascues05}. 

\textit{Extremality of Gaussian states.} 
To prove the optimality of Gaussian collective attacks, we also need a 
very useful theorem, recently proven in \cite{Cirac05}. Let us sketch it here for bipartite states $\rho_{AB}$ that have zero first-order moments.
Let $f$ be a function satisfying the properties
\begin{enumerate}
\item continuity in trace norm: if $\parallel\!\!\rho^{(n)}_{AB}-\rho_{AB}\!\!\parallel_1\rightarrow 0$ when $n\to \infty$, then
$f(\rho^{(n)}_{AB})\rightarrow f(\rho_{AB})$,
\item invariance under local ``Gaussification'' unitaries: 
$f(U_G^{\dagger}\otimes U_G^{\dagger}\; \rho_{AB}^{\otimes N}\; U_G\otimes U_G)=f(\rho_{AB}^{\otimes N})$,
\item strong super-additivity: $f(\rho_{A_{1...N}B_{1...N}})\geq f(\rho_{A_1B_1})+...+f(\rho_{A_NB_N})$ with equality if $\rho_{A_{1...N}B_{1...N}}=\rho_{A_1B_1}\otimes...\otimes \rho_{A_NB_N}$.
\end{enumerate}
Then, for every bipartite state $\rho_{AB}$ with covariance matrix $\gamma_{AB}$, we have that
\begin{equation} \label{extremality}
f(\rho_{AB})\geq f(\rho_{AB}^{G})
\end{equation}
where $\rho_{AB}^{G}$ is the Gaussian state with the same $\gamma_{AB}$. 
The proof can be summarized by
\begin{eqnarray}
f(\rho_{AB})&\stackrel{3}{=}&\frac{1}{N}f(\rho_{AB}^{\otimes N})\stackrel{2}{=}\frac{1}{N}f(\tilde{\rho}_{A_{1...N}B_{1...N}}) 
 \nonumber \\ 
&~& \stackrel{3}{\geq}\frac{1}{N}\sum_{k=1}^{N}f(\tilde{\rho}_{A_kB_k}) \stackrel{1,\star}{\simeq} f(\rho_{AB}^{G})
\end{eqnarray}
where the superscripts label the assumptions used in each step,
while $\tilde{\rho}_{A_{1...N}B_{1...N}} \equiv U_G^{\dagger}\otimes U_G^{\dagger}\; \rho_{AB}^{\otimes N}\; U_G\otimes U_G$.
The $\star$ stands for the use of a central limit result for quantum states (see \cite{Cirac05} for details).  
The Gaussification unitary $U_G$ is a passive operation, 
which can be realized with a network of beam splitters and phase shifters. Importantly for what follows, the $x$ and $p$ quadratures of all $N$ modes are thus not mixed via Gaussification.

\textit{Optimality of Gaussian attacks.} 
The core of our proof now consists in combining this extremality result 
with the entanglement-based version of CV-QKD supplemented with our physical
model of measurement.
In realistic protocols, Alice and Bob do not achieve the Holevo bound, but
only extract the mutual information $I_{ab}=S(a{\rm:}b)$. In contrast, 
Eve is assumed to have no technological limitation, so, by collective attacks, she can attain the Holevo bound $\chi_{aE}=S(a{\rm:}E)$. Then, using our notation, the achievable DR secret key rate reads \cite{Renato05,RenatoPhD},
\begin{equation}
K(\rho_{AB})=S(a{\rm:}b)-S(a{\rm:}E)=S(a|E)-S(a|b).
\label{Kcoll}
\end{equation}
The function $K(\rho_{AB})$ depends on the choice of the measurement done by Alice and Bob (and on the sifting if any), 
but does not depend on the purification of $\rho_{AB}$. We now
will prove that $K(\rho_{AB})$ satisfies the three conditions of the Gaussian extremality theorem. For this, we also need to use the extension of this function over $2N$ modes ($\bar{A}=A_{1...N}$, $\bar{B}=B_{1...N}$), namely 
\begin{equation}
K(\rho_{\bar{A}\bar{B}})=S(\bar{a}{\rm:}\bar{b})-S(\bar{a}{\rm:}E)=S(\bar{a}|E)-S(\bar{a}|\bar{b}).
\label{Kcoll2}
\end{equation}
where Alice (Bob) do the same measurement on her (his) $N$ modes, and Eve 
has the purification of $\rho_{\bar{A}\bar{B}}$. Note that Eq.~(\ref{Kcoll2})
restricts to Eq.~(\ref{Kcoll}) when $N=1$.

\textit{i) Continuity:} If $\parallel\!\!\rho^{(n)}_{\bar{A}\bar{B}}-\rho_{\bar{A}\bar{B}}\!\!\parallel_1\leq \epsilon$,
using Ulhmann's theorem and well-known relations between the fidelity and trace distance \cite{N&C02},
we can find a purification $\ket{\Psi}^{(n)}_{\bar{A}\bar{B}E}$ ($\ket{\Psi}_{\bar{A}\bar{B}E}$) of $\rho^{(n)}_{\bar{A}\bar{B}}$ 
($\rho_{\bar{A}\bar{B}}$) such that
$\parallel\!\!\hat{\Psi}^{(n)}_{\bar{A}\bar{B}E}-\hat{\Psi}_{\bar{A}\bar{B}E}\!\!\parallel_1\leq 2\sqrt{\epsilon}$.
Then, considering that partial trace can only decrease the trace norm \cite{N&C02}, we have 
$\parallel\!\!\rho^{(n)}_{\bar{a}E}-\rho_{\bar{a}E}\!\!\parallel_1\leq 2\sqrt{\epsilon}$
and $\parallel\!\!\rho^{(n)}_{\bar{a}\bar{b}}-\rho_{\bar{a}\bar{b}}\!\!\parallel_1\leq 2\sqrt{\epsilon}$.
Finally, the continuity of von Neumann entropies implies the continuity of $K$.
$\Box$

\begin{figure}[!t!]
\begin{center}
\includegraphics[width=8cm]{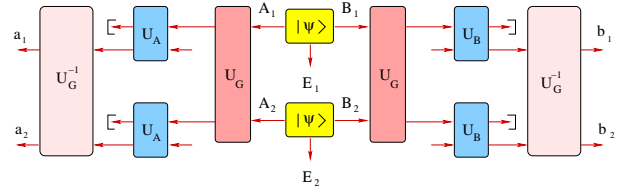}
\end{center}
\caption{Invariance under local ``Gaussification'' unitaries: $U_G$ can be interchanged with the measurement
$U_A$, then $U_G^{-1}$ and $U_G$ cancel each other.}
\label{invariant}
\end{figure}

\textit{ii) Invariance under local Gaussification unitaries:} 
Applying the local Gaussification operation $U_G\otimes U_G$ on the product states $\ket{\psi}_{ABE}^{\otimes N}$
(as shown in Fig.~\ref{invariant} for $N=2$), we obtain the state $\ket{\tilde{\psi}}_{\bar{A}\bar{B}\bar{E}}$. 
After the measurements on Alice's and Bob's sides, the state becomes $\tilde{\rho}_{\bar{a}\bar{b}\bar{E}}$. 
But because the (homodyne or heterodyne) measurement and the Gaussification operation can be interchanged, 
by applying $U_G^{\dagger}\otimes U_G^{\dagger}$ on modes $\bar{a}$ 
and $\bar{b}$ we recover the state $\rho_{abE}^{\otimes N}$, which coincides with the state obtained by directly measuring  
$\ket{\psi}_{ABE}^{\otimes N}$ without Gaussification.
Since the two states $\tilde{\rho}_{\bar{a}\bar{b}}$ and $\rho_{ab}^{\otimes N}$ are related by a
local unitary operation $U_G^{\dagger}\otimes U_G^{\dagger}$ and since the mutual von Neumann entropies appearing
in $K(\rho_{AB})$ are invariant under (any) local unitaries, we obtain the invariance of $K(\rho_{AB})$ under local Gaussification unitaries. $\Box$

\textit{iii) Strong super-additivity:}
We will restrict the proof to two modes on each side, $A_{1,2}$ and $B_{1,2}$, the generalization to $N>2$ being straightforward. 
We have
\begin{equation}
K(\rho_{A_{1,2}B_{1,2}})=S(a_1a_2|E)-S(a_1a_2|b_1b_2)
\label{ssakey}
\end{equation}
where the conditional entropies can be expressed as
\begin{eqnarray}
S(a_1a_2|E)&=&S(a_1|a_2E)+S(a_2|a_1E)+S(a_1{\rm:}b_2|E) \nonumber \\
S(a_1a_2|b_1b_2)&=&S(a_1|b_1b_2)+S(a_2|b_1b_2)-S(a_1{\rm:}a_2|b_1b_2) \nonumber
\end{eqnarray}
As a consequence of the strong sub-additivity of von Neumann entropies,
we obtain the bound
\begin{equation}
K\geq \underbrace{S(a_1|a_2E)-S(a_1|b_1b_2)}_{\geq S(a_1|A_2B_2E)-S(a_1|b_1)}
+\underbrace{S(a_2|a_1E)-S(a_2|b_1b_2)}_{\geq S(a_2|A_1B_1E)-S(a_2|b_2)}
\label{ssakey2}
\end{equation}
(using the fact that conditioning can only decrease the conditional entropy).
The purification of $A_1B_1$ ($A_2B_2$) being
$A_2B_2E$ ($A_1B_1E$), we obtain 
\begin{equation}
K(\rho_{A_{1,2}B_{1,2}})\geq K(\rho_{A_1B_1})+K(\rho_{A_2B_2}).
\label{ssakey3}
\end{equation}
The additivity of $K(\rho_{A_{1,2}B_{1,2}})$ is a straightforward consequence of the additivity of von Neumann entropies. $\Box$

\begin{figure}[!t!]
\begin{center}
\includegraphics[width=8.5cm]{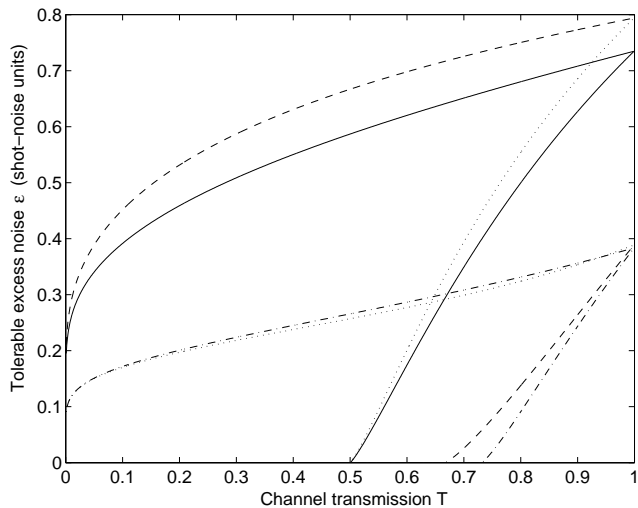}
\end{center}
\caption{Tolerable excess noise $\epsilon$ as a function of the channel transmission $T$ at the limit of an infinite modulation for the four Gaussian protocols: squeezed states + homodyne measurement (solid line),
squeezed states + heterodyne measurement (dashed line),
coherent states + homodyne measurement (dotted line), and
coherent states + heterodyne measurement (dot-dashed line).
The curves vanishing at (or above) $T=0.5$ correspond to DR, whereas those vanishing at $T=0$ refer to RR.}
\label{DR}
\end{figure}

Thus, using Eq.~(\ref{extremality}), we have proved that for all bipartite quantum states $\rho_{AB}$ with covariance matrix $\gamma_{AB}$, one has $K(\rho_{AB})\geq K(\rho^G_{AB})$. This means that $K(\rho^G_{AB})$ is 
a lower bound on the secret key rate for any protocol (even non-Gaussian) and collective attack (including non-Gaussian).
The only requirement for this result to hold is that Alice and Bob
use the second-order moments of the quadratures in order to calculate 
this bound. In particular,
for the Gaussian-modulation protocols 
of \cite{Cerf01,Gross02,Nature03,Ralph04}, 
Eve's optimal attack is a Gaussian attack, in which case the bound is saturated. Note that the above proof concerns DR, see Eq.~(\ref{Kcoll}), but its extension to RR is straightforward: one simply needs to interchange $a\leftrightarrow b$ and $A\leftrightarrow B$. As an illustration, Fig.~\ref{DR} shows the highest tolerable excess noise $\epsilon$ as a function of the line transmission $T$ for the four Gaussian protocols (in DR and RR) and the optimal Gaussian collective attack.

\textit{Coherent attacks} represent the most powerful class of attacks Eve can perform: she let all the signal pulses sent by Alice interact
with a large auxiliary system (quantum computer), which she measures jointly at the end of the classical post-processing. Recently, it has been shown (for discrete-variable QKD) that, under some symmetries of the classical post-processing, the collective attacks are actually as efficient for Eve as the coherent attacks \cite{Renato05,RenatoPhD}. Taking for granted that this proof extends to CV-QKD, we conclude that our optimality proof of Gaussian attacks holds in full generality.

\textit{Realistic implementations} of CV-QKD never achieve the secret key rate 
$K(\rho_{AB})$ because reconciliation protocols are not 100\% efficient.
The actual key rate is
\begin{equation}
K=\beta S(a{\rm:}b)-S(a{\rm:}E)=S(a|E)-\beta S(a|b)-(1-\beta)S(a).
\label{realK}
\end{equation}
where $\beta \in [0,1]$ is the reconciliation efficiency. 
It is easy to prove that Eq.~(\ref{realK}) also satisfies the three conditions of the extremality theorem, so our conclusions remain unchanged. In the special case of $\beta=0$, this means that Eve's accessible information $\chi_{aE}=S(a{\rm:}E)$ is maximized for Gaussian states, so that Gaussian
collective attacks are also optimal in this restricted sense.

\textit{``Quantum'' Bob.} A theoretically interesting -- though probably unrealistic -- situation is the case where Bob reaches the Holevo bound $\chi_{aB}$. This may be done by combining the use of quantum memory with a proper optimal post-processing at Bob's side. 
The ``ultimate'' available secret key rate then reads
\begin{equation}
K=S(a{\rm:}B)-S(a{\rm:}E)=S(a|E)-S(a|B)
\end{equation}
It again satisfies the three above conditions, so it is 
lower bounded by the Gaussian attack.

\textit{Conclusion.}
We have presented a unified analysis of all known QKD protocols based on Gaussian modulation of coherent (or squeezed) states by Alice and homodyne (or heterodyne) detection by Bob, for the DR and RR versions of 
one-way reconcilation. This entanglement-based model of CV-QKD
combined with a physical representation of measurement
gives a very simple way of writing the secret key rates
in terms of mutual von Neumann entropies involving quantum systems (including the preparer and the measurer). Then, exploiting a recent result on the extremality of Gaussian states, we have demonstrated that the optimal collective attack against all these protocols is a Gaussian operation.
It is then sufficient to check the security against Gaussian attacks, which are completely characterized by the covariance matrix $\gamma_{AB}$ estimated by Alice and Bob. This result appears to be quite general as it holds for realistic protocols (with finite reconciliation efficiency) as well as for ideal protocols (where Bob has a quantum memory and extracts the entire accessible information). 
Provided that \cite{Renato05} can be adapted to CV, it even extends to the full unconditional security of CV-QKD against coherent attacks.

Note added: The optimality of Gaussian collective attacks has been independently
proved using different techniques in \cite{unpublished}.

We acknowledge financial support from the EU under projects 
COVAQIAL (FP6-511004) and from the IUAP programme of the Belgian government under grant V-18. R.G.-P. acknowledges support from the Belgian foundation FRIA.

\end{document}